\documentclass[aps,onecolumn,epsfig,graphics,showkeys,floatfix,mathbbm]{revtex4}
\usepackage{amsmath,epsfig}
\usepackage{psfrag,times}
\begin{document}
\title{Scheme for sharing classical information via tripartite entangled states}

\author{Xue Zheng-Yuan\footnote{Emial: zyxue@ahu.edu.cn}, Yi You-Min and
 Cao Zhuo-Liang\footnote{Emial: zlcao@ahu.edu.cn (Corresponding
 author)}}

\affiliation {Key Laboratory of Opto-electronic Information
Acquisition and Manipulation Ministry of Education,\\
and School of Physics and Material Science, Anhui University, Hefei
230039, China}

\begin{abstract}
We investigate schemes for quantum secret sharing and quantum dense
coding via tripartite entangled states. We present a scheme for
sharing classical information via entanglement swapping using two
tripartite entangled GHZ states. In order to throw light upon the
security affairs of the quantum dense coding protocol, we also
suggest a secure quantum dense coding scheme via W state in analogy
with the theory of sharing information among involved users.
\end{abstract}


\maketitle

\section{Introduction}
Quantum entanglement is regarded as a key resource for many tasks in
quantum information processing and quantum communication. Quantum
secret sharing (QSS)\cite{sharing} and quantum dense coding
(QDC)\cite{dense} are two of its most striking applications. QSS is
a process securely distributing private key among three or multiply
parties. If and only if they are in the cooperation with each other,
they can decode the secret message. Meanwhile, if one of them is
dishonest, the honest guy may keep the dishonest one from doing any
damage. There are three main applications of QSS: distributing a
private key among many parties, sharing a classical secret directly
and sharing quantum information. QDC is a process of sending two
classical bits (cbits) of information from a sender (Alice) to a
remote receiver (Bob) by sending only a single qubit. It works in
the following way. Initially, Alice and Bob share a maximally
entangled state. The first step is an encoding process where Alice
performs one of the four local operations on her qubit. Then she
sends the qubit to Bob. The last step is a decoding process. After
Bob has received the qubit, he can identify the local operation of
Alice by using only local operations. Recently, due to their
promising applications in quantum communication, QSS and QDC attract
more and more public attention both theoretically and
experimentally.

Quantum entanglement is a novel result of the superposition
principle in quantum mechanics, and it attracts overwhelming
attention in the newborn subject of quantum
information.\cite{sharing,dense,teleport,1} It is well known that
multipartite qubits can be entangled in different inequivalent ways,
for a tripartite entangled quantum system, it falls into two types
of irreducible entanglements,\cite{entangle} that is, GHZ and W
types state. The motivation of classifying entangled states is that,
if the entanglement of two states is equivalent, then the two states
can be used to perform the same task, although the probability of
successfully performing the task may depend on the degree of
entanglement of the state. GHZ state is a well-established qualified
candidate for QSS, we present here a scheme for QSS via entanglement
swapping with two tripartite entangled GHZ states in Section 2. But,
in the branch of QSS, most schemes utilize the GHZ-type state, and
completely neglect the counterpart type. W-type state is also a
promising candidate for quantum communication schemes and other
tasks in quantum information processing. In order to throw light
upon the security affairs of the quantum dense coding protocol, we
also suggest a secure QDC scheme via W state in Section 3. We refer
to the \emph{secure QDC} as a process securely distributing
information via QDC among many parties in a way only when they are
in cooperation with each other  they can read the distributed
information. The secret message is imposed by local unitary
operations and split into two users via a tripartite entangled W
state. Thus the scheme may be regarded as a combination of QSS and
QDC. This paper ends with a conclusion in Section 4.

\section{QSS via entanglement swapping}

Entanglement swapping\cite{6} is a method of enabling one to
entangle two quantum systems that do not have direct interaction
with each other. Based on it, many applications in quantum
information have been found. Entanglement swapping is also used in
QSS protocols.\cite{7, 8} Here, we present an alternative scheme for
QSS via two tripartite entangled GHZ states. Our scheme, based on
identifying only Bell state and the tripartite quantum channel, can
be checked simultaneously by using the tripartite entangled GHZ
state.

Suppose that Alice wants to send secret information to a distant
agent Bob. As she does not know whether he is honest or not, she
makes the information shared by two users (\textit{i.e}. Bob and
Charlie). If and only if they are in collaboration of each other,
both users can read the information, furthermore, individual users
each could not do any damage to the process. Here, we assume that
the communication over a classical channel is insecure, which means
we can't use the simplest method of teleportation\cite{teleport} to
distribute the information. Of course, one could also fulfill the
task by using standard quantum cryptography, but, on average, it
requires more resources and measurements.\cite{resource} Assume that
Alice initially possesses two tripartite entangled qubits and they
are both prepared in the GHZ-type entangled state
\begin{eqnarray}
\label{abc}
|\psi\rangle_{1,2,3}=1/\sqrt{2}(|000\rangle+|111\rangle)_{1,2,3},
|\psi\rangle_{4,5,6}=1/\sqrt{2}(|000\rangle+|111\rangle)_{4,5,6},
\end{eqnarray}
where qubits 1 and 4 belong to Alice, and qubits 2, 5 and 3, 6 are
sent to Bob and Charlie, respectively. After she confirms that Bob
and Charlie both have received their qubits, the QSS process acts as
follows:

(1) Security checking. In terms of security checking, Refs. [1, 10]
have already proved that the nonlocal correlation with a tripartite
entangled state can perfectly detect any eavesdropping or dishonest
in the process. Only when they confirm by the checking result that
there is no eavesdropper existing in the channel, can they proceed
to encod the secret message, or will the QSS process  be aborted and
start a new round of security checking.

(2) Secret encoding. The secret encoding process is achieved by
local operations on one of the qubits on Alice's side, which is
similar to the encoding process of QDC.\cite{dense} Initially, the
combined state of the 6 qubits is
\begin{equation}
|\psi\rangle=1/2(|eee\rangle+|ggg\rangle)_{1,2,3}(|eee\rangle+i|ggg\rangle)_{4,5,6}.
\end{equation}
Alice performs randomly one of the following four local operations
 \{I, $\sigma^{x}$,
i$\sigma^{y}$, $\sigma^{z}$\} on one of her qubits (representing
two-bit classical information), \textit{e.g}. on qubit 1, which
leads the initial state to
\begin{subequations}
\label{ew}
\begin{eqnarray}
I_{1}|\psi\rangle&=&1/2(|000\rangle+|111\rangle)_{1,2,3}(|000\rangle+|111\rangle)_{4,5,6}\nonumber\\
&=&\sqrt{2}/4[|\Phi^{+}\rangle_{1,4}(|\Phi_{2,5}^{+}\rangle|\Phi_{3,6}^{+}\rangle+|\Phi_{2,5}^{-}\rangle|\Phi_{3,6}^{-}\rangle)
+|\Phi_{1,4}^{-}\rangle(|\Phi_{2,5}^{+}\rangle|\Phi_{3,6}^{-}\rangle+|\Phi_{2,5}^{-}\rangle|\Phi_{3,6}^{+}\rangle)\nonumber\\
&+&|\Psi_{1,4}^{+}\rangle(|\Psi_{2,5}^{+}\rangle|\Psi_{3,6}^{+}\rangle+|\Psi_{2,5}^{-}\rangle|\Psi_{3,6}^{-}\rangle)
-|\Psi_{1,4}^{-}\rangle(|\Psi_{2,5}^{+}\rangle|\Psi_{3,6}^{-}\rangle+|\Psi_{2,5}^{-}\rangle|\Psi_{3,6}^{+}\rangle)],
\end{eqnarray}
\begin{eqnarray}
\sigma^{x}_{1}|\psi\rangle&=&1/2(|100\rangle+|011\rangle)_{1,2,3}(|000\rangle+|111\rangle)_{4,5,6}\nonumber\\
&=&\sqrt{2}/4[(|\Psi_{2,5}^{+}\rangle|\Psi_{3,6}^{+}\rangle+|\Psi_{2,5}^{-}\rangle|\Psi_{3,6}^{-}\rangle)
-|\Phi_{1,4}^{-}\rangle(|\Psi_{2,5}^{+}\rangle|\Psi_{3,6}^{-}\rangle+|\Psi_{2,5}^{-}\rangle|\Psi_{3,6}^{+}\rangle)\nonumber\\
&+&|\Psi_{1,4}^{+}\rangle(|\Phi_{2,5}^{+}\rangle|\Phi_{3,6}^{+}\rangle+|\Phi_{2,5}^{-}\rangle|\Phi_{3,6}^{-}\rangle)
-|\Psi_{1,4}^{-}\rangle(|\Phi_{2,5}^{+}\rangle|\Phi_{3,6}^{-}\rangle+|\Phi_{2,5}^{-}\rangle|\Phi_{3,6}^{+}\rangle)],
\end{eqnarray}
\begin{eqnarray}
i\sigma^{y}_{1}|\psi\rangle&=&1/2(|100\rangle-|011\rangle)_{1,2,3}(|000\rangle+|111\rangle)_{4,5,6}\nonumber\\
&=&\sqrt{2}/4[|\Phi_{1,4}^{+}\rangle(|\Psi_{2,5}^{+}\rangle|\Psi_{3,6}^{-}\rangle+|\Psi_{2,5}^{-}\rangle|\Psi_{3,6}^{+}\rangle)
-|\Phi_{1,4}^{-}\rangle(|\Psi_{2,5}^{+}\rangle|\Psi_{3,6}^{+}\rangle+|\Psi_{2,5}^{-}\rangle|\Psi_{3,6}^{-}\rangle)\nonumber\\
&-&|\Psi_{1,4}^{+}\rangle(|\Psi_{2,5}^{+}\rangle|\Psi_{3,6}^{-}\rangle+|\Psi_{2,5}^{-}\rangle|\Psi_{3,6}^{+}\rangle)
-|\Psi_{1,4}^{-}\rangle(|\Psi_{2,5}^{+}\rangle|\Psi_{3,6}^{+}\rangle+|\Psi_{2,5}^{-}\rangle|\Psi_{3,6}^{-}\rangle)],
\end{eqnarray}
\begin{eqnarray}
\sigma^{z}_{1}|\psi\rangle&=&1/2(|000\rangle-|111\rangle)_{1,2,3}(|000\rangle+|111\rangle)_{4,5,6}\nonumber\\
&=&\sqrt{2}/4[|\Phi_{1,4}^{+}\rangle(|\Phi_{2,5}^{+}\rangle|\Phi_{3,6}^{-}\rangle+|\Phi_{2,5}^{-}\rangle|\Phi_{3,6}^{+}\rangle)
+|\Phi_{1,4}^{-}\rangle(|\Phi_{2,5}^{+}\rangle|\Phi_{3,6}^{+}\rangle+|\Phi_{2,5}^{-}\rangle|\Phi_{3,6}^{-}\rangle)\nonumber\\
&+&|\Psi_{1,4}^{+}\rangle(|\Phi_{2,5}^{+}\rangle|\Phi_{3,6}^{-}\rangle+|\Phi_{2,5}^{-}\rangle|\Phi_{3,6}^{+}\rangle)
+|\Psi_{1,4}^{-}\rangle(|\Phi_{2,5}^{+}\rangle|\Phi_{3,6}^{+}\rangle+|\Phi_{2,5}^{-}\rangle|\Phi_{3,6}^{-}\rangle)],
\end{eqnarray}
\end{subequations} where
$|\Phi^{\pm}\rangle_{k,j}=1/\sqrt{2}(|00\rangle\pm|11\rangle)_{k,j}$
and $
|\Psi^{\pm}\rangle_{k,j}=1/\sqrt{2}(|01\rangle\pm|10\rangle)_{k,j}$
are the four Bell states and $(k,j)$=(1,4), (2,5) or (3,6).

(3) Secret extracting. Obviously, one can see that there is an
explicit correspondence between the local operation and the
Bell-state measurement outcomes of the three users from Eq.
(\ref{ew}). That is, if Alice informs Bob and Charlie of her
measurement result to through a public channel, then the the two
users in cooperation can read the secret message. For example, if
the declared measurement of Alice is $|\Phi_{1,4}^{-}\rangle$, the
measurement of Bob is $|\Phi_{2,5}^{-}\rangle$ and he informs
Charlie of his result, then Charlie knows the local operation of
Alice from his local measurement, \textit{i.e.}
$|\Phi_{3,6}^{-}\rangle\rightarrow\sigma^{z}_{1}$ and
$|\Phi_{3,6}^{+}\rangle\rightarrow I_{1}$.

In this way, we have presented a three-party QSS scheme based on
quantum entanglement swapping and identification of Bell states.
With measurements, they can identify Alice's operation on the qubit
1. So the two users share two bits of classical information from
Alice. This is very reasonable, in the protocols of Ref.
\cite{sharing}, they adopt one tripartite entangled GHZ state and
thus can only shares one bit information between Alice and Bob.

Now we turn to discuss the security of our scheme.

(1) If there is an eavesdropper who has been able to entangle an
ancilla with the quantum channel, and later he can measure the
ancilla to gain information about the measurements from the legal
users. However, Hillery \textit{et al}. \cite{sharing} show that if
this entanglement does not introduce any errors into the procedure,
then the state of the system is a product of the GHZ state and the
ancilla, which means that the eavesdropper could gain nothing about
the measurements on the triplet from observing the ancilla.

(2) If one of the users is the eavesdropper, say Bob, who wants to
obtain Alice's information without the cooperation which the third
party and without being detected. If he can always reveal his
measurements after Alice and Charlie, then he can succeed in
cheating the other two users.\cite{sharing2} But, this can be easily
avoided, Alice can requires the two users to declare their
measurements in turns or even in random order.

(3) Bob can also obtain the qubit that Alice sends to Charlie, and
sends Charlie a qubit that he has prepared before hand in order to
read Alice's message without Charlie's help. As Bob does not know
Alice's measurements before hand, thus the qubit he sent to Charlie
is not in the correct quantum state. By checking the measurements
with Alice publicly, the eavesdropping behaviour of Bob can be
detected.

\section{Secure QDC via W state}

QDC, despite of its novel capacity in terms of sending classical
information, should be very deliberately used for the sake of
security in the process, thst is, the receiver Bob can always use
the information willingly. Now, the question arises: does any secure
QDC scheme exist? We note that QSS is likely to play an important
role in protecting secret quantum information. In this section, we
present a secure QDC scheme via W state.

Assume the three parties, \textit{i.e}. Alice Bob and Charlie, to
share a tripartite entangled that was prepared in the following W
state in advance
\begin{equation}
\label{w0}
|\psi\rangle_{1,2,3}=1/\sqrt{3}(|001\rangle+|010\rangle+|100\rangle)_{1,2,3},
\end{equation}
where particles 1, 2 and 3 belong to Alice, Bob and Charlie,
respectively.

1. Information encoding. Alice performs one of the four local
operations $\{\textit{I}, \sigma^{x},i\sigma^{y}, \sigma^{z}\}$  on
her qubit, which represent two-bit classical information. These
operations will transform the state (\ref{w0}) into
\begin{subequations}
\label{w1}
\begin{equation}
|\psi\rangle_{1}=1/\sqrt{3}(|001\rangle+|010\rangle+|100\rangle)_{1,2,3},
\end{equation}
\begin{equation}
|\psi\rangle_{2}=1/\sqrt{3}(|101\rangle+|110\rangle+|000\rangle)_{1,2,3},
\end{equation}
\begin{equation}
|\psi\rangle_{3}=1/\sqrt{3}(|101\rangle+|110\rangle-|000\rangle)_{1,2,3},
\end{equation}
\begin{equation}
|\psi\rangle_{4}=1/\sqrt{3}(|001\rangle+|010\rangle-|100\rangle)_{1,2,3}.
\end{equation}
\end{subequations}
Now the information is encoded into the pure entangled state, which
is shared among the three parties.

2. Qubit transmission. Alice sends her qubit to one of the two
receivers (say Bob), we will latter find out the party that Alice
sends her qubit to is not arbitrary. After a party receives the
qubit, he will have a high probability of successful cheat compared
with the one who has not in the QDC procedure. So, Alice would send
her qubit to the party, which is less likely to cheat in the
process.

3. Information extracting. Assume that Bob selected above has
received Alice's qubit, then the state in Eq. (\ref{w1}) can be
rewritten as
\begin{subequations}
\label{w2}
\begin{equation}
|\psi\rangle_{1}=1/\sqrt{6}(|\Phi^{+}_{1,2}\rangle+|\Phi^{-}_{1,2}\rangle)|1\rangle_{3}+\sqrt{2/3}|\Psi^{+}_{1,2}\rangle|0\rangle_{3},
\end{equation}
\begin{equation}
|\psi\rangle_{2}=1/\sqrt{6}(|\Psi^{+}_{1,2}\rangle+|\Psi^{-}_{1,2}\rangle)|1\rangle_{3}+\sqrt{2/3}|\Phi^{+}_{1,2}\rangle|0\rangle_{3},
\end{equation}
\begin{equation}
|\psi\rangle_{3}=1/\sqrt{6}(|\Psi^{+}_{1,2}\rangle+|\Psi^{-}_{1,2}\rangle)|1\rangle_{3}+\sqrt{2/3}|\Phi^{-}_{1,2}\rangle|0\rangle_{3},
\end{equation}
\begin{equation}
|\psi\rangle_{4}=1/\sqrt{6}(|\Phi^{+}_{1,2}\rangle+|\Phi^{-}_{1,2}\rangle)|1\rangle_{3}+\sqrt{2/3}|\Psi^{-}_{1,2}\rangle|0\rangle_{3},
\end{equation}
\end{subequations}
where $|\Phi^{\pm}_{1,2}\rangle$ and $|\Psi^{\pm}_{1,2}\rangle$ are
the four Bell states of particles 1 and 2. Obviously, one can see
from Eq. (\ref{w2}) that if the result of Bob's measured is
$|1\rangle_{3}$ then the QDC process fails. If Bob's measured result
is $|0\rangle_{3}$ then there is an explicit correspondence between
Alice's operation and the measured results of the two receivers.
Thus if they are in cooperation with each other, both of them can
obtain the information. So, our scheme is a probabilistic one, with
the successful probability $P=2/3$.

But if they do not choose to cooperate with each other, neither of
the two users could obtain the information by local operation in a
deterministic manner. Now, let us turn to the case that they do not
choose to cooperate with each other. If Charlie lies to Bob, Bob
also has a probability of $2/3$  to obtain the correct information,
so the successful cheat probability of Charlie is $1/3$. Conversely,
Charlie only has a probability of $1/4$ to have the correct
information, so the successful cheat probability of Bob is $3/4$.
This is the point that we have mentioned in Step 2, that is, he who
received Alice's particle has a higher probability of successful
cheat compared with the party who has not (see Eq. (\ref{w2})).

We also note that the scheme can be generalized to the case of
multi-users providing Alice possesses a multipartite entangled
state. Suppose that she has a (\textit{N}+1)-qubit entangled state,
qubits $2,3,\cdot\cdot\cdot(\textit{N}+1)$ are to \textit{N} users,
respectively. After she confirms that each of the users has received
a qubit, she then operates one of the four local measurements on the
qubit 1. After that, the two-cbit information is encoded into the
(\textit{N}+1)-qubit entangled state. Later, she sends her qubit to
one of the rest \textit{N} users. Again, he who has received the
qubit will have a high probability of successful cheat compared with
the rest (\textit{N}-1) users. Only in the cooperation with all the
rest users,  can one obtain Alice's information. In this way, we set
up a network for secure QDC via QSS.

\section{Conclusion}

We have investigated a QSS sharing protocol via entanglement
swapping using two tripartite entangled GHZ states. If and only if
when they are in the cooperation with each other, they can read the
original information. Any attempt to obtain the complete information
of the state without the cooperation with the third party cannot be
succeed in a deterministic way. We also presented a scheme for
secure QDC with a tripartite entangled W state. The scheme is
probabilistic but secure one. If and only if when they are in the
cooperation with each other, they can identify Alice's local
operational. The two schemes based on identifying only Bell state,
thus require less demands for experimental demonstration.

\acknowledgements Project supported by Anhui Provincial Natural
Science Foundation (Grant No 03042401), the Key Program of the
Education Department of Anhui Province (Grant Nos. 2002kj029zd and
2006kj070A), and the Talent Foundation of Anhui University. One of
the authors, Xue Zheng-Yuan, is also supported by the Postgraduate
Innovation Research Plan from Anhui university.

\end{document}